\documentclass[11pt,a4paper]{article}
\usepackage{amsfonts}
\usepackage{amsmath,amssymb}
\usepackage{epsfig,graphicx}

\begin{document}

\begin{center}
\bigskip \bigskip

{\LARGE \textbf{Vector-Field Domain Walls }}

\bigskip \bigskip \bigskip \bigskip

\textbf{J.L.~Chkareuli}$^{a,b}$, \textbf{Archil Kobakhidze}$^{a,c}$\textbf{\
and Raymond R. Volkas}$^{c}$

\bigskip

$^{a}$\textit{E. Andronikashvili Institute of Physics, 0177 Tbilisi, Georgia 
}

$^{b}$\textit{I. Chavchavadze State University, \textit{0162} Tbilisi,
Georgia\ \vspace{0pt}\\[0pt]
}

$^{c}$\textit{School of Physics, The University of Melbourne,Victoria 3010,
Australia}\\[0pt]

\bigskip {\scriptsize E-mail: j.chkareuli@iliauni.edu.ge ,
archilk@unimelb.edu.au, raymondv@unimelb.edu.au \\[0pt]
\bigskip }\bigskip \bigskip \bigskip \bigskip \bigskip

\textbf{Abstract}
\end{center}

\bigskip

We argue that spontaneous Lorentz violation may generally lead to metastable
domain walls related to the simultaneous violation of some accompanying
discrete symmetries. Remarkably, such domain wall solutions exist for
space-like Lorentz violation and do not exist for the time-like violation.
Because a preferred space direction is spontaneously induced, these domain
walls have no planar symmetry and produce a peculiar static gravitational
field at small distances, while their long-distance gravity appears the same
as for regular scalar-field walls. Some possible applications of
vector-field domain walls are briefly discussed. 
\thispagestyle{empty}\newpage

\section{Gauge fields as vector Nambu-Goldstone bosons}

Relativistic invariance and local gauge symmetries are cornerstones of our
current understanding of elementary particles and their interactions.
Elementary particles as quanta of their corresponding fields are classified
through irreducible representations of the Poincar\'{e} group, while the
local gauge symmetries prescribe their dynamics. This theoretical picture is
most successfully realized within the celebrated Standard Model of quarks
and leptons and their fundamental strong, weak and electromagnetic
interactions.

Nevertheless, it is conceivable that local gauge symmetries and the
associated masslessness of gauge bosons might have a completely different
origin, being in essence dynamical rather than due to a fundamental
principle. This point of view is partially motivated by the peculiarities of
local gauge symmetries themselves, which, unlike global symmetries,
represent redundancies of the description of a theory rather than being
\textquotedblleft true\textquotedblright\ symmetries. In addition, the
dynamical origin of massless particle excitations is very well understood in
terms of spontaneously broken global symmetries. Based on these
observations, the origin of massless gauge fields as the vector
Nambu-Goldstone bosons appearing due to spontaneous Lorentz invariance
violation (SLIV) \cite{Bjorken:1963vg} has gained new impetus \cite%
{Chkareuli:2001xe}-\cite{Chkareuli:2007da} in recent years.\footnote{%
Independently of the problem of the origin of local symmetries, Lorentz
violation in itself has attracted considerable attention as an interesting
phenomenological possibility which may be probed in direct Lorentz
non-invariant extensions of quantum electrodynamics (QED) and the Standard
Model \cite{jakiw}-\cite{glashow}.}

While the first models realizing the SLIV conjecture were based on the four
fermion interaction (where the photon was expected to appear as a
fermion-antifermion composite state \cite{Bjorken:1963vg}), the simplest
model for SLIV is in fact given by a conventional QED type Lagrangian
extended by an arbitrary vector field potential energy. For the minimal
polynomial containing only the vector field bilinear and quadrilinear terms
one comes to the Lagrangian 
\begin{equation}
\mathcal{L}=-\frac{1}{4}F_{ab}F^{ab}-\frac{\lambda }{4}\left(
A_{a}A^{a}-n^{2}v^{2}\right) ^{2},\text{ }  \label{1}
\end{equation}%
where $n_{a}$ ($a=0,1,2,3$) is a properly-oriented unit Lorentz vector, $%
n^{2}=n_{a}n^{a}=\pm 1$, while $\lambda $ and $v^{2}$ are, respectively,
dimensionless and mass-squared dimensional positive parameters. This
potential means in fact that the vector field $A_{a}$ develops a constant
background value $\langle A_{a}\rangle =n_{a}v$\ and Lorentz symmetry $%
SO(1,3)$ breaks down to $SO(3)$ or $SO(1,2)$ depending on whether $n_{a}$ is
time-like ($n_{a}^{2}>0$) or space-like ($n_{a}^{2}<0$). Expanding the
vector field around this vacuum configuration, 
\begin{equation}
A_{a}(x)=n_{a}(v+\phi )+\mathcal{A}_{a}(x)~,\text{ \ }n_{a}\mathcal{A}^{a}=0
\label{2}
\end{equation}%
one finds that the $\mathcal{A}_{a}$ field components, which are orthogonal
to the Lorentz violating direction $n_{a}$, describe a massless vector
Nambu-Goldstone boson, while the $\phi $ field corresponds to a Higgs mode.

This minimal polynomial QED extension, Eq.\ (\ref{1}), is sometimes referred
to as the \textquotedblleft bumblebee\textquotedblright\ model (see paper 
\cite{Bluhm:2004ep} and references therein). Its nonlinear version \cite%
{nambu} with a directly imposed vector field constraint $%
A_{a}^{2}=n^{2}v^{2} $ (which appears virtually in the limit $\lambda
\rightarrow \infty $ \ from the potential (1)) was also intensively
discussed in the literature (see paper \cite{Chkareuli:2007da} and
references therein). Actually, both of these models are equivalent in the
infrared energy domain, where the Higgs mode is considered infinitely
massive, and amount to gauge invariant QED taken in the axial gauge, as was
shown in tree \cite{nambu} and one-loop \cite{Azatov:2005wv} approximations.
\ This axial gauge, $n_{a}\mathcal{A}^{a}=0$, singles out the pure
Goldstonic modes in the vector field as per Eq.\ (\ref{2}).

\section{Vector field domain walls in flat spacetime}

We show here that systems described by Lagrangians that are extensions of type (\ref{1})
may possess topologically stable\footnote{We shall discuss what we mean by `` topologically stable''
in this context in more detail below.  But in short:  it is well-known that SLIV theories
do not have stable ``vacua'', since the Hamiltonians are not bounded
from below.  Since ``vacua'', which are more accurately to be termed ``local minima'', 
are used as the boundary conditions for domain wall solutions, those solutions can never be
absolutely stable.  Nevertheless, local minima will be metastable.  Topological (meta-)stability
then means that the local minima used as boundary conditions for domain wall solutions
form a disconnected manifold,
just as in the usual case of absolutely-stable topological domain walls in scalar 
field theories.  See also
\cite{bazeia} for a discussion of topological defects in scalar field theories 
with explicit Lorentz invariance violation.}
domain wall configurations. Remarkably,
such domain wall solutions exist for space-like Lorentz violation and do not
exist for time-like violation. 

One may at first suspect that the topological stability of
such domain wall solutions may stem from the fact that the discrete $Z_{2}$
symmetry $A_{a}(x)\rightarrow -A_{a}(x)$ of the Lagrangian (\ref{1}) lies
outside the connected part of the $SO(1,3)$ Lorentz symmetry group and, when
minimal couplings to matter fields $\psi (x)$ are included, is nothing but
the charge conjugation invariance 
\begin{equation}
C:\text{ }A_{a}(x)\rightarrow -A_{a}(x),~\psi (x)\rightarrow \psi ^{\mathrm{c%
}}(x).
\end{equation}%
The nonzero VEVs $\langle A_{a}\rangle =\pm n_{a}v$ spontaneously break this
symmetry. We may thus search for a vector-field domain wall solution which
asymptotes to the two $Z_{2}$ degenerate vacua, $+vn_{a}$ and $-vn_{a}$,
along some spatial direction. \ \ 

However, the model, as it stands in Eq.(\ref{1}), does not yet have topologically
stable walls. The point is that the potential terms in the model (\ref{1}) have an
extra accidental symmetry $SO(1,3)^{\prime }$ which rotates vector fields
like the Lorentz symmetry while leaving the space-time coordinates
untransformed. This $SO(1,3)^{\prime }$ is in fact a generic internal
symmetry of the potential in the model (\ref{1}) or in any other QED polynomial
extension. As a result, the vacuum state four-vector of type $+vn_{a}$ can
be transformed into the vector $-vn_{a}$ by a continuous $SO(1,3)^{\prime }$
rotation, so the system does not possess disconnected vacua.
For pure geometrical reasons,
this argument holds just for space-like domain wall solutions ($n^{2}=-1$)
and would not be applicable to time-like domain walls ($n^{2}=1$), if such solutions
might appear through spontaneous Lorentz violation. However, as we show
below, only the space-like domain wall solutions specifically appear in the
model of type (\ref{1}) so one might think that they all are unstable.

To achieve stability, the model can be extended, as we shall see shortly, by
introducing a second vector field $B_{a}$ into the model (\ref{1}). There
naturally appears, apart from the discrete $Z_2$ symmetry ($%
A_{a}\rightarrow -A_{a}$, $B_{a}\rightarrow -B_{a}$), also the interchange
symmetry $A\leftrightarrow B$ in the properly arranged Lagrangian model of
the $A$ and $B$ fields.  We shall show that the spontaneous breakdown of
this interchange discrete symmetry may provide stability for vector-field
domain walls. 

We consider first the one-vector field case to illustrate some
generic features of vector-field domain walls, and then proceed to the two-vector field model.

\subsection{One-vector field model}

The general domain wall solution in flat spacetime can be searched for using
the simple ansatz, 
\begin{equation}
A_{a}(x)=n_{a}V(\mathfrak{n}\cdot x)~,  \label{3}
\end{equation}%
where $\mathfrak{n}_{b}$ is another constant unit four-vector, $\mathfrak{n}%
\cdot x=\mathfrak{n}_{b}x^{b}$. The equation of motion following from (\ref%
{1}) then reads 
\begin{equation}
\left[ \mathfrak{n}^{2}n_{a}-(\mathfrak{n}\cdot n)\mathfrak{n}_{a}\right]
V^{\prime \prime }(\mathfrak{n}\cdot x)-\lambda n^{2}n_{a}\left( V^{2}(%
\mathfrak{n}\cdot x)-v^{2}\right) V(\mathfrak{n}\cdot x)=0~,  \label{4}
\end{equation}%
where primes denote the differentiation with respect to the argument of a
function. As one can immediately see multiplying the equation (\ref{4}) by
vector $\mathfrak{n}^{a}$ that \ static domain wall solutions appear only
for the pure space-like spontaneous Lorentz violation with orthogonal $\ n$
and $\mathfrak{n}$ directions, 
\begin{equation}
\ n^{2}=\mathfrak{n}^{2}=-1,\text{ \ \ }n_{a}\mathfrak{n}^{a}=0\text{ }.
\label{4'}
\end{equation}
Then the equation (\ref{4}) for $V$ reduces to the familiar equation for the
scalar domain wall: 
\begin{equation}
V^{\prime \prime }-\lambda V\left( V^{2}-v^{2}\right) =0~.  \label{5}
\end{equation}%
The solution for the above equation is well known: 
\begin{equation}
V(\mathfrak{n}\cdot x)=v\tanh \left[ m\text{ }(\mathfrak{n}\cdot x)\right] ~,
\label{6}
\end{equation}%
where $m=\sqrt{\lambda /2}v$.

Consider for definiteness a domain wall extending along the $z$-direction,
i.e. $\mathfrak{n}_{a}=(0,0,0,1)$, and centered at $z=0$ with the Lorentz
violating vacua taken in the $y$-direction, i.e. $n_{a}=(0,0,1,0)$: 
\begin{equation}
A_{2}=v\tanh \left( mz\right) ~,~~A_{0}=A_{1}=A_{3}=0~.  \label{7}
\end{equation}%
Observe that due to the Lorentz non-invariant VEVs, this configuration does
not have an $x-y$ planar symmetry, contrary to the case of the ordinary
scalar field domain wall \cite{Vilenkin:1981zs}, \cite{Vilenkin:1984hy}, 
\cite{Ipser:1983db}. This important difference is reflected in the
energy-momentum tensor of our domain wall configuration (\ref{7}): 
\begin{eqnarray}
T^{ab} &\equiv &-F_{~c}^{a}F^{bc}-\lambda A^{a}A^{b}\left(
A_{c}A^{c}-n^{2}v^{2}\right) +\frac{1}{4}\eta ^{ab}\left[ F_{cd}F^{cd}+%
\lambda \left( A_{c}A^{c}-n^{2}v^{2}\right) ^{2}\right]  \notag \\
&=&\frac{m^{2}v^{2}}{\cosh ^{4}\left( mz\right) }diag\left[ 1,-1,-2\sinh^2
\left( mz\right) ,0\right] ~.  \label{8}
\end{eqnarray}%
Note the fact that the preferred Lorentz-violating direction in the plane of
the vector domain wall implies that $T^{11}\neq T^{22}$ in contrast to a
standard scalar domain wall, though its surface energy density, 
\begin{equation}
\sigma =\int_{-\infty }^{+\infty }T_{0}^{0}dz=\frac{4}{3}mv^{2}~,  \label{8a}
\end{equation}%
is the same as for a scalar wall, as directly follows from Eq.(\ref{8}).

However, as was argued above, this vector domain wall solution is
unstable since the vacuum $(0,0,+v,0)$ in this model can be continuously
changed to the vacuum $(0,0,-v,0)$ by a proper rotation in the $x-y$ plane.
So, one has to think about some extensions of the model among which the
two-vector field system seems to be the simplest possibility.

\subsection{Two-vector field model}

The most general Lagrangian of two vector fields $A$ and $B$ possessing the
independent discrete symmetries,
\begin{eqnarray}
& A \to -A,\quad B \to B,& \nonumber\\
& A \to A,\quad B \to -B,&
\label{Z2minus}
\end{eqnarray}
together with the
interchange symmetry $I$,
\begin{equation}
A\leftrightarrow B,
\label{Z2I} 
\end{equation}
may be written in the following simple form,
\begin{eqnarray}
\mathcal{L}_{(A,B)} &=&-\frac{1}{4}F_{A}^{2}-\frac{1}{4}F_{B}^{2}-\frac{\mu
^{2}}{2}(A^{2}+B^{2})-\frac{\alpha }{4}\left(A^{4}+B^{4}\right)   \notag \\
&&-\frac{\beta }{2}(A^{2}B^{2})-\text{ }\frac{\gamma }{2}(AB)^{2},  \label{l2}
\end{eqnarray}%
in a self-evident notation for contractions of $A$ and $B$ fields and their
field-strength tensors ($A^{2}=A_{a}A^{a}$, $AB=A_{a}B^{a}$, $A^{4}=(A_{a}A^{a})^{2}
$, $F_{A}^{2}=F_{Aab}F_{A}^{ab}$ etc.) and positive dimensionless coupling
constants $\alpha$, $\beta$  and $\gamma $. The sign of the
mass parameter, $\mu ^{2}/2$, has been chosen so as to have just 
space-like Lorentz violation in the model,
as was argued in the above. 

Note that, as in the one-field model (\ref{1}), 
the potential in $\mathcal{L}_{(A,B)}$ also possesses
an accidental symmetry $SO(1,3)_{A}^{\prime }\times
SO(1,3)_{B}^{\prime }$ which is broken to a ``diagonal'' subgroup by the
$\gamma$-term in  Eq.(\ref{l2}).  Also, when $\alpha = \beta$, these two terms
combine into $(A^2 + B^2)^2$ which possess an internal $SO(2)$ also
exhibited by the mass and kinetic terms.  But this continuous symmetry is explicitly
broken by the $\gamma$-term, and also if $\alpha \neq \beta$. The
non-existence of this $SO(2)$ will be important for establishing 
domain wall topological stability.

One can readily confirm that this potential has the following extrema.
The first extremum has both the $A$ and $B$ fields identically equal to zero,
but it is arranged to be a local maximum, providing $\mu^2 > 0$.
Second, there is the ``symmetrical'' extremum given by the following
expectation values of the fields: 
\begin{equation}
\langle A_{a}\rangle = \langle B_{a}\rangle = 
n_{a}\sqrt{\frac{\mu ^{2}}{\alpha +\beta +\gamma}},
\label{v1}
\end{equation}%
where $n_a$ is a spacelike unit vector, $n_a^2= -1$.
If this were the vacuum,\footnote{Recall that by ``vacuum'' we really mean metastable local minimum.} 
it would spontaneously break the discrete symmetries
of Eq.(\ref{Z2minus}) while leaving the interchange symmetry $I$ of Eq.(\ref{Z2I}) exact.

There are also ``asymmetric'' extrema.  One class is given by
\begin{equation}
\langle A_{a}\rangle =n_{a}v,\text{ }\langle B_{a}\rangle =0\text{\ \
\ \ }\quad {\rm with}\quad v\equiv \sqrt{\mu ^{2}/\alpha },  
\label{v2}
\end{equation}%
which is degenerate with the $I$-related configuration
\begin{equation}
\langle A_{a}\rangle =0,\text{ }\langle B_{a}\rangle =n_{a}v.
\label{v'2}
\end{equation}%
A vacuum of this type breaks $I$ and one of the $Z_2$ symmetries of Eq.(\ref{Z2minus}).
The other class is given by
\begin{equation}
\langle A_{a}\rangle =n_{a}v,\text{ }\langle B_{a}\rangle = n_b v\text{\ \
\ \ }\quad {\rm with}\quad v\equiv \sqrt{\frac{\mu ^{2}}{\alpha+\beta}},  
\label{v3}
\end{equation}%
where $n_a^2 = n_b^2 = -1$ and $n_a \cdot n_b = 0$.  It is degenerate with its $I$-symmetry
correlate.

Note that the vacua of Eqs.(\ref{v1}) and (\ref{v3}) are unsuitable for domain-wall stability
since all configurations of those types
form connected manifolds. For example, there are two-field analogues of the
solution we derived in the one-vector field model.  One such configuration is
\begin{equation}
A_{a}(x)=n_{a}v\tanh [m(\mathfrak{n}\cdot x)],~B_{a}(x)=0
\end{equation}%
and another is its $I$-symmetry partner. 
However, for the parameter space region where Eqs.(\ref{v2}) and (\ref{v'2})
are the vacua, a qualitatively different wall solution exists, related to the
spontaneous breaking of the interchange symmetry $I$ 
rather than the reflection symmtries of Eq.(\ref{Z2minus}). We find that the extremum (\ref{v2}) (and (\ref{v'2})) is indeed realized as a local minimum if the coupling constants satisfy the inequalities 
\begin{equation}
0<\alpha \leq \frac{\beta + \gamma}{2}~,~~\alpha \leq \beta.
\label{ineq}
\end{equation}
All the other extrema discussed above are local maxima when the inequalities (\ref{ineq}) hold.
Note that in this range of parameters the potential in (14) is not bounded from below. The reason 
is that $(A^2B^2)$ term is neither positive nor negative definite, and thus for $0<\alpha < \beta$ 
one always finds a direction in $(A,B)$ space along which the potential runs to $-\infty$. However, 
it is unreasonable to insist on the boundness of the potential since, as mentioned earlier, the 
total Hamiltonian is known to be unbounded \cite{Bluhm:2008yt}, \cite{Carroll:2009em}. The 
corresponding instabilities are caused by sufficiently strong fluctuations of the Higgs 
modes of vector fields around a vacuum. For small (linear) perturbations (e.g., in the low 
energy regime of the theory), however, stability can be maintained \cite{Carroll:2009em}. 
Thus we only demand that vacua are realized as local minima, and they are stable under the 
linear perturbations. We will confirm shortly below that our domain wall solutions are also 
stable perturbatively against linear perturbations of the Higgs modes.

A prototype for a domain wall solution has the form 
\begin{equation}
A_{a}(x)=n_{a}v(1+\tanh [m(\mathfrak{n}\cdot x)])/2,\text{ \ \ }%
B_{a}(x)=n_{a}v(1-\tanh [m(\mathfrak{n}\cdot x)])/2  \label{ws}
\end{equation}%
which asymptotes to the two $I$-degenerate vacua, 
\begin{equation}
\langle (A_{a},B_{a})\rangle =(vn_{a},0)\ \ \text{and}\ \ (0,vn_{a})
\end{equation}%
along some spatial direction $\mathfrak{n}\cdot x \in (-\infty,+\infty)$. One can
easily see that these vacua cannot be rotated to each other in principle.
The only continuous symmetry that might do it is the $SO(2)$ acting in the (%
$A,B$) space, but this symmetry is explicitly broken in the Lagrangian, as discussed earlier.

The prototype solution above is the simplest analytic 
configuration of this type one can write down using 
the standard hyperbolic tangent function.  In theories involving multiple fields, other
kinds of kink-like functions can also be solutions, though they can usually only be
obtained numerically.  It is often the case that the analytic prototype corresponds
to a particular relationship holding amongst the parameters in the potential, and if
that relationship does not hold then wall solutions still exist but must be computed
numerically.

To establish that wall solutions of the type we want exist, it suffices to
constrain ourselves to the analytic prototype and show that it indeed can be a solution.
Putting the ansatze (\ref{ws}) into the equations of motion of the vector
fields $A_{a}(x)$ and $B_{a}(x)$ one finds that such a solution indeed exists,
with $m=\sqrt{\lambda /2}v=\mu /\sqrt{2}$, in the parameter plane
\begin{equation}
\beta +\gamma =3\alpha
\end{equation}%
This relation is consistent with the potential
stability conditions (\ref{ineq}), providing $\gamma \leq 2 \beta$. 
As emphasized already, for other values of these constants a numerical integration of
the equations of motion is required.

Consider, again as in the one-vector field case, a domain wall extending
along the $z$-direction, i.e. $\mathfrak{n}_{a}=(0,0,0,1)$, and centered at $%
z=0$ with the Lorentz violating vacua taken both in the $y$-direction, i.e. $%
n_{a}=(0,0,1,0)$: 
\begin{equation}
A_{2}(z) = v [ 1 + \tanh (m z) ]/2,\quad B_{2}(z)=v [ 1 - \tanh (m z) ])/2  
\label{sol2}
\end{equation}%
while $A_{0}=A_{1}=A_{3}=0$ and $B_{0}=B_{1}=B_{3}=0$. Equally, they could
be taken in the orthogonal $x$- and $y$-directions, respectively. In any case,
due to the preferred space directions in these vacua, our domain wall
configuration does not possess an $x-y$ planar symmetry and has, as it can
easily be confirmed , the following energy-momentum tensor 
\begin{equation}
T^{ab}=\frac{m^{2}v^{2}}{2\cosh ^{4}\left( mz\right) }diag\left[ 1,-1,-2\sinh^2
\left( mz\right) ,0\right]  \label{ten2}
\end{equation}%
being exactly half of the corresponding tensor in 
the one-vector field case (\ref{8}).

To conclude, we found vacuum configurations in the model where
only one of the two vector fields $A$ or $B$ is condensed, thus producing
three Goldstone modes collected into a photon-like multiplet, while its
counterpart has a mass of order the Lorentz violation scale and decouples
from the low-energy physics. Actually, the mass-squared of the non-condensed
field being from the outset negative (tachyonic) becomes now positive. If,
say, $A$ is condensed (while $B$ is not) the mass of the $B$-field exitations along $n_a$ direction
is given by
\begin{equation}
M_{B_{\|}}=\mu \sqrt{(\beta +\gamma )/\alpha -1}=\mu \sqrt{2},
\end{equation}
while the excitations orthogonal to $n_a$ have the mass,
\begin{equation}
M_{B_{\bot}}=\mu \sqrt{\beta/\alpha -1}.
\end{equation}
So, on one side of the wall one has massless $A$ photons and
massive $B$ bosons, while on the other side obe has massive $A$ bosons
and massless $B$ photons. In this sense, the model contains in
fact only one massless vector field in all possible observational
manifestations.

\subsection{Linear stability}

Before discussing the gravitational properties of our two-field domain wall 
solutions we would like to show that the wall solution is stable against 
linear perturbations in the Higgs modes, $\phi_A$ and $\phi_B$: 
$A_{\mu}= n_{a}(V_A+\phi_A)+a_{a}$ and $B_{a}= n_{\mu}(V_B+\phi_B)+b_{a}$. 
Here $V_A=v(1+\tanh [m(\mathfrak{n}\cdot x)])/2$ and $V_B=v(1-\tanh [m(\mathfrak{n}\cdot x)])/2$ 
correspond to the background solutions in (25). We plug the above expansion into the equations of motion obtained from the Lagrangian (14). Then, using the equation of motions for the background solutions $V_A$ and $V_B$ and keeping only the terms linear in $\phi_A$ and $\phi_B$, we obtain the 
linearized equations of motion for the Higgs modes:
\begin{eqnarray}
\left[\Box +(n\cdot \partial)^2\right]\phi_A =(\mu^2-3\alpha(V_A^2+V_B^2))\phi_A - 6\alpha V_AV_B\phi_B~, \\
\left[\Box +(n\cdot \partial)^2\right]\phi_B =(\mu^2-3\alpha(V_A^2+V_B^2))\phi_B - 6\alpha V_AV_B\phi_A~, 
\end{eqnarray}
The stability can be easily established by looking at linear combinations: 
$\phi_{+}=\phi_A+\phi_B$ and $\phi_{-}=\phi_A-\phi_B$. Namely, it can be 
straightforwardly obtained from (27) and (28) that the mode $\phi_+$ is 
simply a massive $(m_+=\sqrt{2}\mu)$ free field with the (Lorentz-violating) dispersion relation $\omega^2=2\mu^2+\vec{k}^2-(\vec n\cdot \vec k)^2>0$.  The solution for the the mode $\phi_{-}$ can be written as $\phi_{-}=N_{-}{\rm e}^{-i\omega+ik_xx+ik_yy}f(z)$. For $f(z)$ we obtain, 
\begin{equation}
f''(z)+\left[-\omega^2+k_x^2+\mu^2(1+3\tanh^2(mz))\right]f(z)=0
\end{equation}
This equation is essentially the same as the corresponding equation obtained in 
the case of the the usual one-field scalar domain wall. Hence, 
the domain wall solutions are perturbatively stable against linear Higgs mode 
perturbations, despite the fact that the total Hamiltonian is not bounded from below. 

A more generic analysis of the linear perturbations (including those of 
Goldstonic modes $a_{a}$ and $b_{a}$) is a complicated problem 
and can not be handled analytically. However, we expect that the Goldstonic  
perturbations do not induce instabilities due to the topological 
reasons. Indeed, due to the topological charge conservation, the domain wall configuration (21) can "decay" by emitting $a_{a}$ and $b_{a}$ quanta only into a configuration with the same topological charge and lower surface energy. However, it is not difficult to check that any domain wall configuration which interpolates between $I$-degenerate vacua, $(n_{a}'v, 0)$ and $(0, n_{a}'v)$, have the same energy density $T^{00}$ (25). Therefore, the only source of the instability of domain walls is the metastability of the vacua (16,17), and this instability shows up at non-linear level.  Thus we argue that the wall solutions are indeed metastable, though we acknowledge that this has not been explicitly checked through linear stability analysis.

\section{Gravitational field of a vector-field domain wall}

Next we are interested in the gravitational properties of vector-field
domain walls. Let us start by considering the weak gravity approximation, 
\begin{equation}
g_{\mu \nu }(x)=\eta _{\mu \nu }+h_{\mu \nu }(x)~.  \label{a}
\end{equation}%
(where the tensors now have \textquotedblleft curved\textquotedblright\
indices $\mu ,\nu ...$ rather than the previous \textquotedblleft
flat\textquotedblright\ indices $a,b,...$). In this approximation we can use
the flat spacetime solution for the vector domain wall (\ref{sol2}) and the
associated energy-momentum tensor (\ref{ten2}). In the harmonic gauge, $%
\partial ^{\mu }\left( h_{\mu \nu }-1/2\eta _{\mu \nu }h\right) =0$, the
Einstein equations are, 
\begin{equation}
h_{\mu \nu }^{\prime \prime }(z)=\frac{2}{M_{\mathrm{P}}^{2}}\left( T_{\mu
\nu }-1/2\eta _{\mu \nu }T\right) ~,  \label{b}
\end{equation}%
where $T=T_{\mu }^{\mu }$ is the trace of the energy-momentum tensor. The
reflection ($z\rightarrow -z$) symmetric solution to the above equations is 
\begin{eqnarray}
h_{00} &=&-h_{11}=-\frac{v^{2}}{6M_{\mathrm{P}}^{2}}\left[ 2\ln (\cosh
(mx))+\cosh ^{-2}(mz)\right] ,  \notag \\
h_{22} &=&\frac{v^{2}}{3M_{\mathrm{P}}^{2}}\left[ \ln \left( \cosh
(mz)\right) -\cosh ^{-2}(mz)\right]  \label{c} \\
h_{33} &=&\frac{v^{2}}{M_{\mathrm{P}}^{2}}\left[ \ln \left( \cosh
(mz)\right) \right] ~.  \notag
\end{eqnarray}

One can see a peculiar static gravitational field of the vector wall at
small distances. Interestingly, one of the $h$ components, namely $h_{33},$ is
vanishing at small distances $mz \ll 1$. At the same time all $h$ components
grow linearly with $|z|$ away from the wall's center. Thus, strictly
speaking, the weak field approximation is not valid at large $|z|$, so one
has to find a solution in the full nonlinear theory. As is well known \cite%
{Vilenkin:1984hy}, \cite{Ipser:1983db}, planar symmetric spacetime metrics
(in the case of scalar-field domain walls) in Einstein gravity without a
cosmological constant are necessarily time-dependent. We find below that the
same appears in our case as well despite the planar symmetry being broken for
vector-field domain walls.

In principle, we have to simultaneously solve the Einstein equations and the
equations of motion for the vector field in the spacetime given by an {\it a priori}
unknown metric. Analytic solutions are difficult, if not impossible, to
obtain. To proceed further we will thus consider the thin wall approximation
which is sufficient for our purpose because we are interested here in how
the metric behaves away from the wall (where the above linear approximation
is not valid). Indeed, the finite-thickness domain-wall solutions must
approach thin-wall-limit solution in regions distant from the wall, where
the\textquotedblleft microscopic\textquotedblright\ details of the wall
structure are not essential. In the thin-wall limit , $m\rightarrow \infty $
such that $mv^{2}=\mathrm{const.}$, the energy-momentum tensor (\ref{ten2})
takes the form, 
\begin{equation}
T_{\mu \nu }=\sigma \delta (z)\mathrm{diag}\left[ 1,-1,-1,0\right] ~,
\label{13a}
\end{equation}%
which coincides with the energy-momentum tensor for the scalar field domain
wall. Therefore, we eventually come to the same solution\cite%
{Vilenkin:1984hy}, \cite{Ipser:1983db} for metric 
\begin{equation}
ds^{2}\equiv g_{\mu \nu }dx^{\mu }dx^{\nu }=(1-k\left\vert z\right\vert
)^{2}[dt^{2}-\mathrm{e}^{2kt}(dx^{2}+dy^{2})-dz^{2}]~.  \label{9}
\end{equation}%
where $k=2\pi G_{N}\sigma .$

We would like to conclude this section with the following remark concerning
the thin wall approximation, comparing a limiting procedure from
polynomial theory (which causes Lorentz violation) to the case of standard QED.
Consider for simplicity the one-vector model given by the 
Lagrangian (\ref{1}). As we have pointed out in the end of the Section 1, 
in the limit $\lambda \rightarrow \infty $ such that $v$ remains finite, 
which is different from
the thin-wall limit considered immediately above, the potential in (\ref{1})
reduces to the gauge field constraint $A_{\mu }^{2}=n^{2}v^{2}$, and this
model is in fact equivalent to a gauge invariant $U(1)$ theory which may be
identified with QED taken in the nonlinear gauge. This is also a type of
thin-wall limit, since $m\rightarrow \infty $, but in this case the energy
density $\sigma $ diverges and thus the wall, even if it might be stable in
this generic model, would be infinitely massive and decoupled from
low-energy physics. Therefore, standard QED, though it may be considered
as a low-energy approximation for a general polynomial vector field theory (%
\ref{1}), is fundamentally free from domain-wall type solutions.

\section{Cosmological evolution of vector-field domain walls}

The vector-field domain walls described above can be produced in the early
universe through the Kibble mechanism \cite{Kibble:1976sj}. Ordinary (scalar
field) domain walls evolve cosmologically as per $\rho \sim 1/a$ in the
non-relativistic limit, where $a$ is the cosmological scale factor. They
therefore dominate the energy density of the universe from very early times
unless the corresponding discrete symmetry breaking scale $v$ is less than $%
\sim 100$ MeV. In fact, it must be less even than $\sim 1$ $MeV$ if the
anisotropy induced by the walls on the cosmic microwave background radiation
is to be below experimental limits\footnote{%
In the context of the inflationary scenario, the domain wall problem exists
if the universe gets reheated enough to allow the production of domain walls
after inflation, i.e. $T_{\mathrm{reheat.}}>v$. Otherwise, the domain walls
are either inflated away from the visible universe, or they simply are not
produced.}.

The cosmological evolution of the vector-field domain walls seems to largely
follow to the same scenario as the regular scalar field walls. However,
there may be some difference as well. A full analysis of this problem is
beyond the scope of this paper, but we do wish to make some simple
observations.

One reason for a difference between scalar and vector domain walls is
the violation of Lorentz invariance on the domain wall surface.
Heuristically, this can be understood as follows. Because of Lorentz
symmetry breaking on the surface of a vector-field wall [$SO(1,2)\rightarrow SO(1,1)$],
tangential motion along the Lorentz-violating direction becomes also
relevant for the dynamical evolution of a domain wall, while the motion
along the Lorentz-preserving direction is still unobservable. As a result,
the vector-field wall may effectively behave in a different way from a scalar-field wall.

Another reason is the generic dynamical instability in 
polynomial theories \cite{Bluhm:2008yt}, \cite{Carroll:2009em} already
discussed. In general, such an
instability can only be removed by removing the Lorentz Higgs mode thus
going to pure QED in an axial or non-linear gauge. For a very heavy Higgs, 
this instability may be vanishingly
small. In general, however, this instability will make the vector domain walls
unstable, even though the ``vacua'' serving as boundary conditions are
topologically disconnected.  This may be welcome phenomenologically:
absolutely stable vector-field walls would be as
phenomenologically unacceptable as the usual stable scalar-field walls.
Long-lived quasi-stable walls (whose
lifetime is controlled by the heavy mass of Lorentz Higgs mode) may be well
suited to the course of cosmological evolution.

Due to the above reasons, the phenomenological bounds for
vector walls should be differetn from those for ordinary walls, and may be weaker. 
Detailed numerical
modelling of a network of vector-field domain walls would probably be
necessary to compute the precise bounds, including from the generation of
microwave background anisotropy.

\section{Conclusion and outlook}

We have argued that spontaneous Lorentz-invariance violation
leads to domain wall solutions related to the simultaneous violation of the
accompanying discrete symmetries, which may be charge-conjugation and/or
interchange symmetry.  Depending on the specific model, such a configuration
may be unstable or metastable.

As was illustrated by the example of the simple
one-vector field model, such domain wall solutions can exist for
space-like Lorentz violation and do not exist for time-like violations.
Though the one-vector field model leads to an unstable wall, the two-vector field
extension seems to also be of physical interest. There naturally appears,
apart from the discrete $Z_{2}$ symmetries ($A_{a}\rightarrow -A_{a}$, $%
B_{a}\rightarrow -B_{a}$), the interchange symmetry $A\leftrightarrow B$ in
an appropriate Lagrangian model of the $A$ and $B$ fields. This additional discrete
symmetry provides a mechanism to achieve metastability for vector-field domain walls. We found a
metastable vacuum configuration in the model where only one of the two vector fields
condenses on each side of the wall, thus producing Goldstone modes collected into a
photon-like multiplet while its counterpart acquires the Lorentz violation
scale order mass and decouples from the low-energy physics. Because a
preferred space direction is spontaneously induced, these domain walls have
no planar symmetry and produce a peculiar static gravitational field at
small distances, while their long-distance gravity appears the same as for
regular scalar-field walls. Cosmological bounds on these domain walls may be
weaker than for the usual scalar walls, but a more detailed analysis is
required before this can be quantified.

The vector-field domain walls are especially interesting if this QED type
model is further extended to the Standard Model (where the Lorentz-violating
vector field is then taken to be coupled to the hypercharge current rather
than the electromagnetic one), and also to grand unified theories. Because
the discrete symmetry that is spontaneously broken may be the charge
conjugation, one application could be to baryogenesis, since particles and
antiparticles are expected to behave differently due to their interactions
with a wall. Another application might concern the extension of the model to
higher dimensions and the possibility of trapping of gauge fields (both
Goldstonic and non-Goldstonic) to a $4$-dimensional vector-field domain wall
appearing in the higher dimensional bulk. 
We may return to these interesting points elsewhere.

\paragraph{Acknowledgments}

One of us (J.L.C.) appreciates the warm hospitality shown to him during his
visit to the Theoretical Particle Physics Group at the University of
Melbourne where part of this work was carried out. The visiting scholar
award received from the University of Melbourne is also gratefully
acknowledged. The work of A.K. and R.R.V. was supported in part by the
Australian Research Council. We thank Alexander Vilenkin for interesting
comments.



\end{document}